\documentclass[conference,10pt]{IEEEtran}

\usepackage[T1]{fontenc}

\IEEEoverridecommandlockouts

\usepackage{amsmath,amssymb,amsfonts}
\usepackage{algorithmic}
\usepackage{graphicx}
\usepackage{textcomp}
\usepackage{xcolor}
\usepackage[capitalise]{cleveref}
\usepackage{siunitx}
\usepackage{subfig}
\usepackage{xcolor}
\usepackage{array}
\usepackage{multicol}
\usepackage{multirow}
\usepackage{tabularx}
\usepackage{tabularray}
\usepackage{colortbl}

\definecolor{blue}{rgb}{0.0, 0, 0}
\usepackage[backend=biber, sorting=none, bibstyle=ieee,citestyle=numeric-comp]{biblatex}
\small{
    \bibliography{references.bib}
}    
\DefineBibliographyStrings{english}{%
  mathesis = {MSc. Thesis},
}

\begin{document}

\title{Evaluation of the Energy Consumption of a Mobile Robotic Platform for Sustainable Wireless Networks}

\author{
	\IEEEauthorblockN{
	Diogo Ferreira, André Coelho, Rui Campos}
	\IEEEauthorblockA{INESC TEC and Faculdade de Engenharia, Universidade do Porto, Portugal\\
    up202006808@edu.fe.up.pt, andre.f.coelho@inesctec.pt, rui.l.campos@inesctec.pt}
}%

\maketitle

\begin{abstract}
The proliferation of wireless devices requires flexible network infrastructures to meet the increasing Quality of Service (QoS) requirements. Mobile Robotic Platforms (MRPs) acting as mobile communications cells are a promising solution to provide on-demand wireless connectivity in dynamic networking scenarios. However, the energy consumption of MRPs is a challenge that must be considered to maximize the availability of the wireless networks created.

The main contribution of this paper is the experimental evaluation of the energy consumption of an MRP acting as a mobile communications cell. The evaluation considers different actions performed by a real MRP, demonstrating that energy consumption varies significantly with the type of action performed. The results obtained pave the way for optimizing MRP movement in dynamic networking scenarios, maximizing wireless network's availability while minimizing the MRP energy consumption.
\end{abstract}


\section{Introduction\label{sec:Introduction}}

\begin{figure}
  \begin{center}
    \includegraphics[width=1\columnwidth]{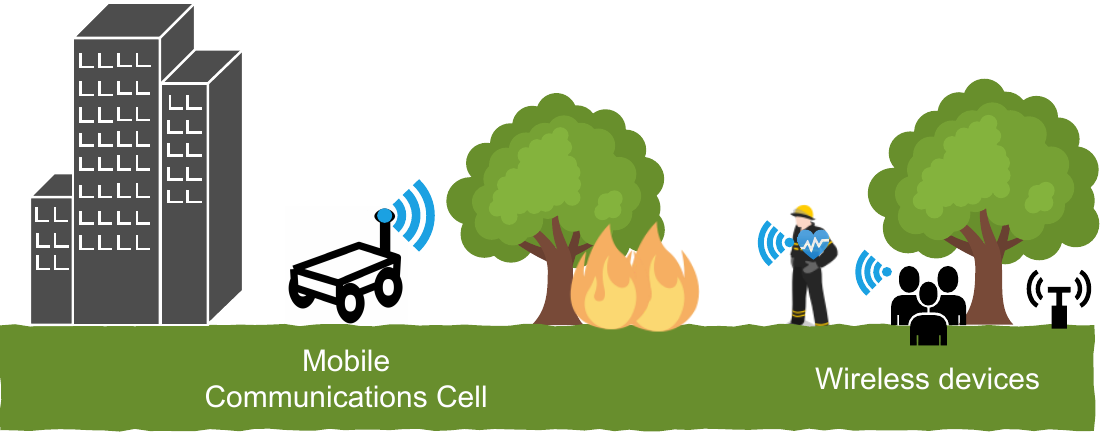}
    \caption{Mobile communications cell providing on-demand connectivity to wireless devices in a disaster management scenario.}
    \label{fig:Rede_a_pedido}
  \end{center}
\end{figure}

In recent years, society's need for wireless network connectivity has grown exponentially. Simultaneously, the 6G paradigm is emerging, paving the way for widespread usage of immersive applications and wireless devices connected to each other, including sensors, actuators, wearable devices, remotely controlled robots, and autonomous vehicles, operating anytime, anywhere \cite{Wang2023, Tataria2021, Romeo2020, Vermesan2020}. With the increasing number of wireless devices and the emergence of online services and applications, reinforcing wireless network infrastructures to meet the increased Quality of Service (QoS) requirements has become necessary. This has motivated the use of communications cells carried by Mobile Robotic Platforms (MRPs) to restore and enhance wireless connectivity on-demand in locations with limited coverage and capacity \cite{Liu2021}. This is especially relevant in temporary events, such as disaster management scenarios depicted in \cref{fig:Rede_a_pedido}, which occur unexpectedly and for short periods, making the deployment of permanent wireless network infrastructures impracticable and non-cost-effective. 

A critical challenge when deploying mobile communications cells compared with fixed networks is the MRPs' endurance. Since MRPs are not permanently connected to the electrical grid, they rely on their on-board batteries, which may deplete quickly. For this reason, MRPs typically need periodic battery recharges; this limits the availability of the wireless networks created. MRPs acting as mobile communications cells consume energy for two main tasks: communications and movement. While the energy spent for communications is typically minimal, the energy spent for movement is substantial. Previous research on Unmanned Aerial Vehicles (UAVs) concluded that power consumption is not uniform: it is lower for low-speed values than for hovering, and it increases with speed. Therefore, hovering is not the most power-efficient state for UAVs \cite{Rodrigues2022_Joint-Energy-Performance}. Still, research works are focused on reducing energy consumption for communications rather than movement \cite{Ludovica2023}. Characterizing the energy consumption associated with each type of MRP movement will enable selecting the most energy-efficient movements for deploying mobile communications cells. This aspect is especially relevant when MRPs move to adjust their positioning and improve wireless connectivity offered, while accommodating factors such as the mobility of wireless devices and dynamic obstacles that may potentially obstruct signal propagation.

The main contribution of this paper is the experimental evaluation of the energy consumption of an MRP acting as a mobile communications cell for providing on-demand wireless connectivity in dynamic networking scenarios. This evaluation is especially relevant for maximizing battery longevity per cycle and improving the availability of the wireless network created by MRPs. Ultimately, the results presented in this paper will enable the selection of energy-efficient movements for MRPs to allow for minimal energy consumption and maximum endurance.

The rest of this paper is organized as follows. 
\cref{sec:SystemDesign} details the system specification. 
\cref{sec:SystemValidation} explains the experimental energy consumption evaluation carried out, including the obtained results.
\cref{sec:Discussion} discusses the results obtained and the main limitations of the performance evaluation conducted.
Finally, \cref{sec:Conclusions} presents the main conclusions and directions for future work.

\section{System Specification\label{sec:SystemDesign}} 

The MRP used in this research work was Unitree Go1 Edu \cite{UnitreeRobotics}.
It is a quadruped robot developed for research and development of autonomous systems in transportation and human interaction. This MRP can be controlled by a remote-controlled wireless command or using an Application Programming Interface (API), enabling control of the MRP movement via Hypertext Transfer Protocol (HTTP). This MRP weighs approximately 12 kg and has a payload capacity up to 5 kg, making it suitable to carry a communications node, such as a Wi-Fi Access Point or cellular Base Station.

\begin{figure}[!ht]
  \begin{center}
    \includegraphics[width=0.4\columnwidth]{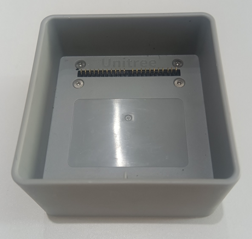}
    \caption{MRP's battery charging base.}
    \label{fig:charging_base}
  \end{center}
\end{figure}

\begin{figure}
  \begin{center}
    \includegraphics[width=0.9\columnwidth]{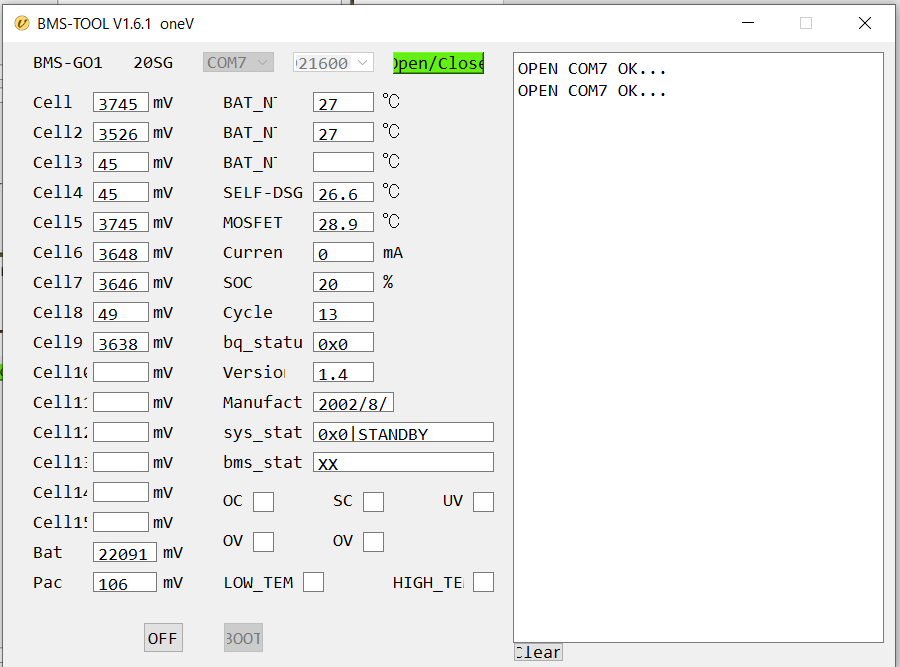}
    \caption{Application's interface used to perform experimental measurements on the MRP's battery.}
    \label{fig:APP}
  \end{center}
\end{figure}

The MRP's battery was a lithium-ion battery with a rated capacity of 4500 mAh, a rated voltage of 21.6 V, and a limit charge voltage of 25.2 V. The battery is charged using the charging base depicted in \cref{fig:charging_base}. The charging base also measures the voltage of each cell and the state of charge of the battery, expressed by \cref{eq:SoC}. The measurements were made through a software application provided by Unitree, depicted in \cref{fig:APP}. 
It provides the voltage of each of the nine battery cells, total voltage, battery temperature, charging cycle and state of charge.

\begin{equation}
State~of~Charge~(\%) = \frac{Remaining~Capacity}{Total~Capacity}\times100
\label{eq:SoC}
\end{equation}

\section{Experimental Energy Consumption Evaluation\label{sec:SystemValidation}} 

\begin{figure}
\centering
\begin{minipage}{.5\columnwidth}
  \centering
  \includegraphics[width=0.9\columnwidth]{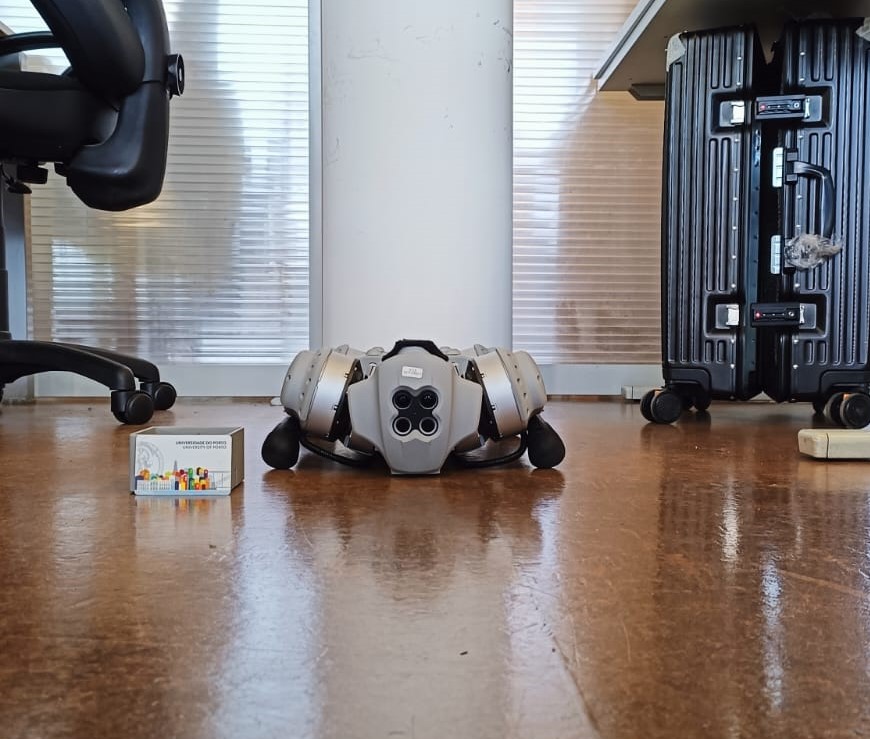}
  \captionof{figure}{MRP laying down.}
  \label{fig:lay_down}
\end{minipage}%
\begin{minipage}{.5\columnwidth}
  \centering
  \includegraphics[width=0.9\columnwidth]{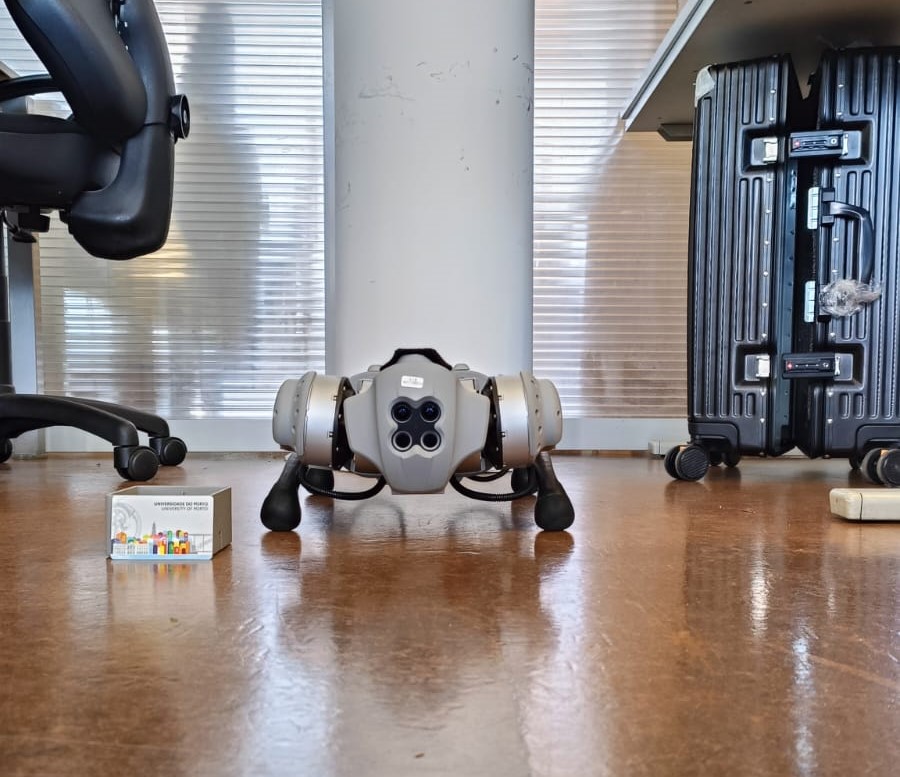}
  \captionof{figure}{MRP crouching.}
  \label{fig:crouching}
\end{minipage}
\end{figure}

\begin{figure}
  \begin{center}
    \includegraphics[width=0.5\columnwidth]{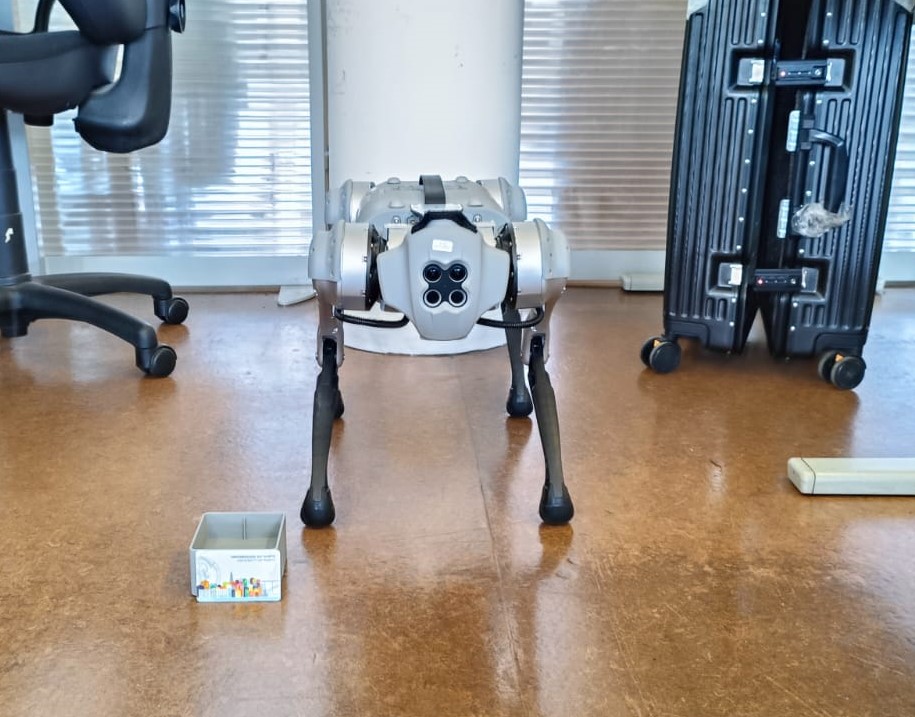}
    \caption{MRP standing.}
    \label{fig:Standing}
  \end{center}
\end{figure}

In order to characterize the energy consumption of the MRP, four reference actions were assessed:
\begin{enumerate}
\item \textbf{Lay down position}. As shown in \cref{fig:lay_down}, this is the most stable position, with the entire MRP in contact with the ground. It does not exert any significant strain on the motors, requiring mainly power for the operation of the onboard computing unit.

\item \textbf{Crouch position}. As depicted in \cref{fig:crouching}, the MRP has its limbs tucked in, requiring some effort from the motors to counteract the force of gravity.

\item \textbf{Standing position}. As illustrated in \cref{fig:Standing}, the MRP's limbs are fully extended. The motors must work to counteract gravity and maintain the MRP's balance.

\item \textbf{Forward walk}. This is the most demanding action among the four studied. It requires the most effort from the motors to move forward and maintain an upright position (cf. \cref{fig:Standing}). Additionally, the motors must work to keep the MRP balanced while moving at a constant speed of 0.4 m/s.
\end{enumerate}

The energy consumption measurements were carried out by following these steps:

\begin{enumerate}
\item \textbf{Installation}. The battery was placed on the charging base and a Universal Serial Bus (USB) cable was connected between the charging base and a computer.

\item \textbf{Software Execution}. Custom software was used to extract battery state measurements from the Unitree application, following these steps:

\begin{itemize}
    \item \textbf{Initiation}. The Unitree application was launched and the appropriate configuration for the measurements was selected.
    \item \textbf{Configuration}. MRP's battery power button was pressed once, followed by the \emph{Enter} key in the terminal.
    \item \textbf{Data Collection}. The custom software was used to retrieve the values from \emph{Cell} to \emph{Cell9} (cf. \cref{fig:APP}), as well as \emph{Cycle}, and state of charge (\emph{SoC}) fields.
\end{itemize}

\item \textbf{Action Performance}. A given action (laying down, crouching, standing, or forward walk) was executed by the MRP and its duration was measured.

\item \textbf{Iteration}. For ensuring statistical variability, the entire process was repeated multiple times until significant battery discharge prevented further iterations.
\end{enumerate}

\begin{figure}
    \centering
    \subfloat[\centering Energy consumption in laying down position.]{{\includegraphics[width=0.80\columnwidth]{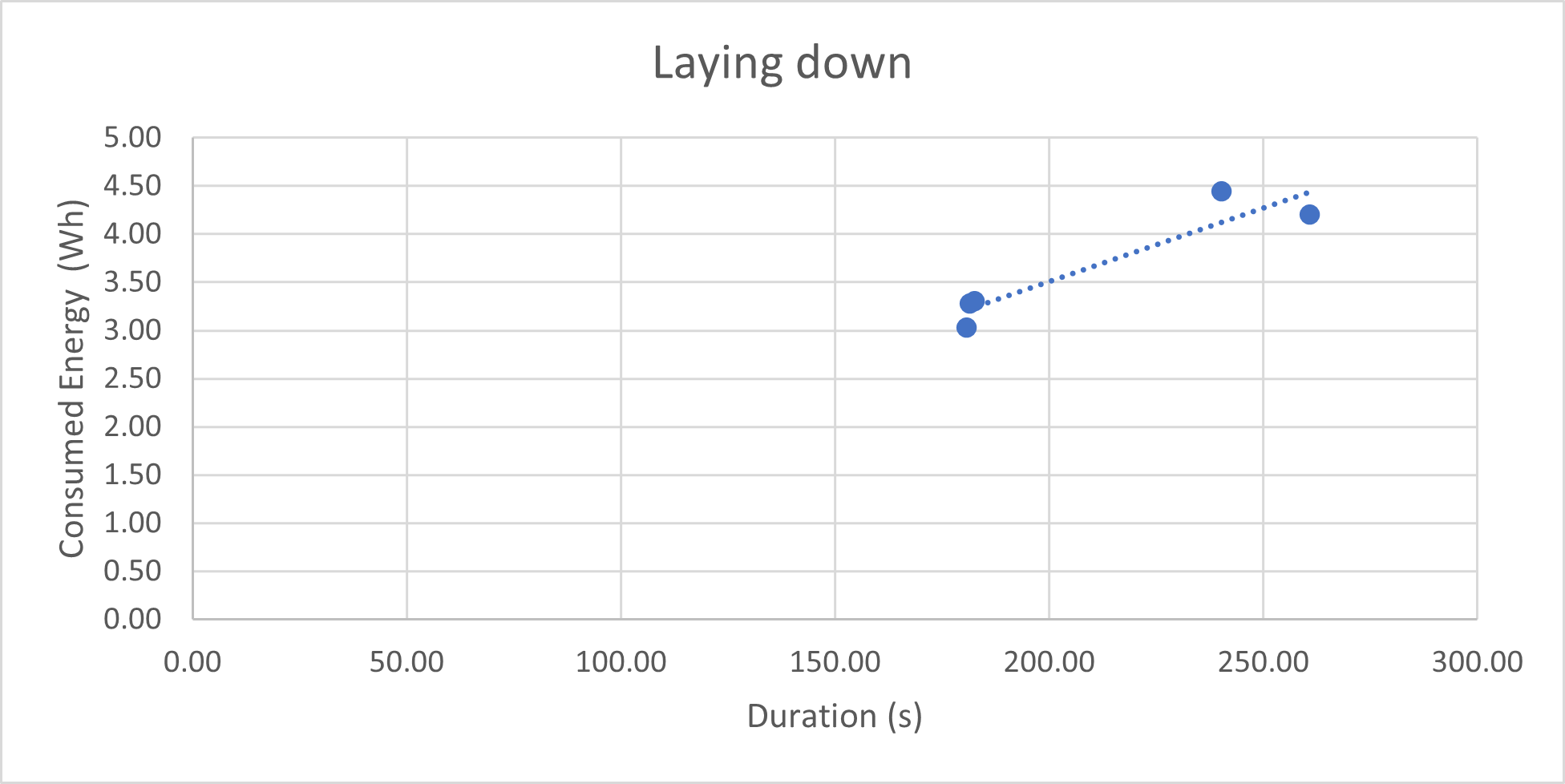}\label{fig:energy_layingdown}}}%
    \qquad
    \subfloat[\centering Energy consumption in crouching position.]{{\includegraphics[width=0.80\columnwidth]{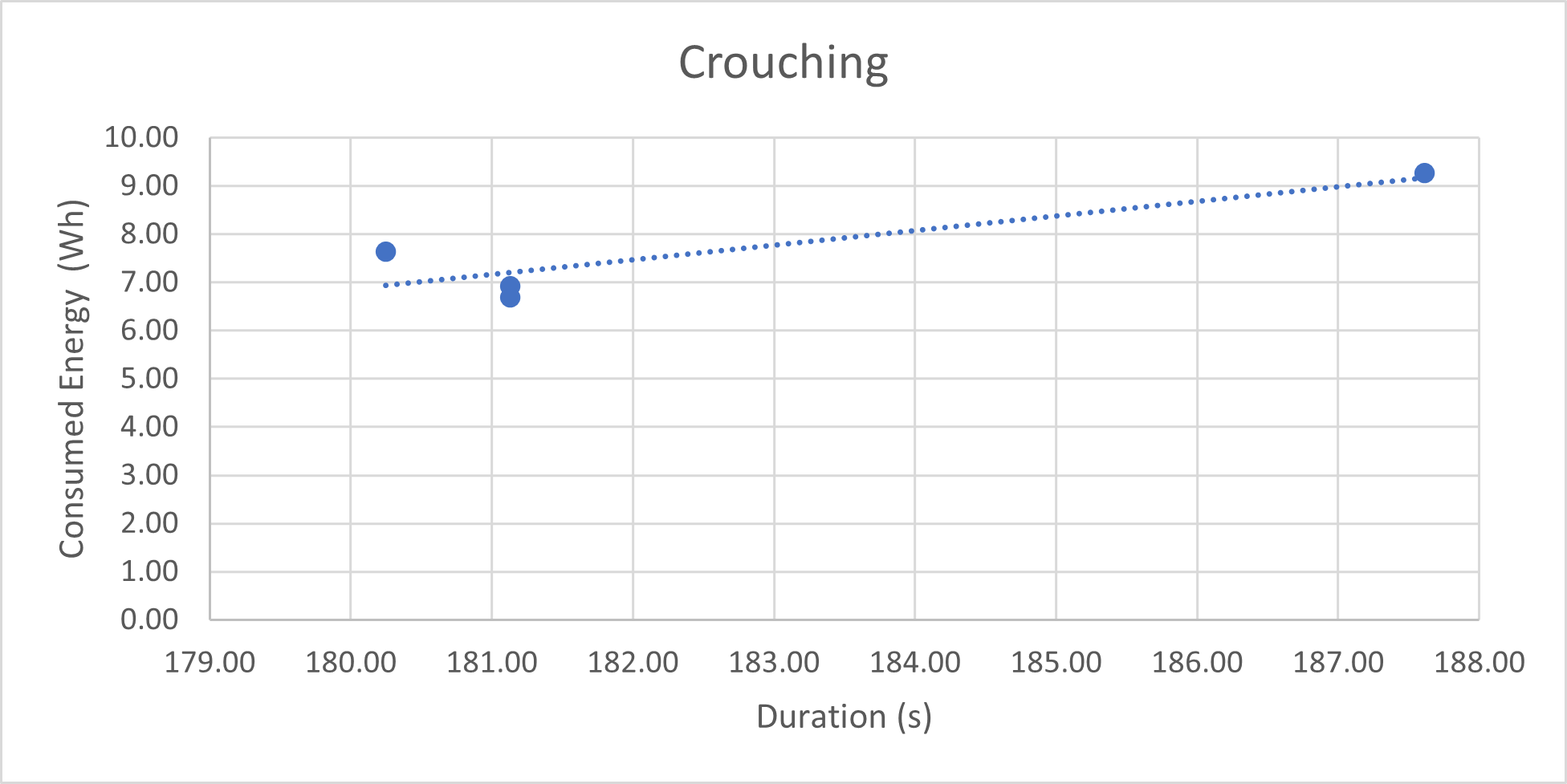}\label{fig:energy_crouching}}}%
    \qquad
    \subfloat[\centering Energy consumption in standing position.]{{\includegraphics[width=0.80\columnwidth]{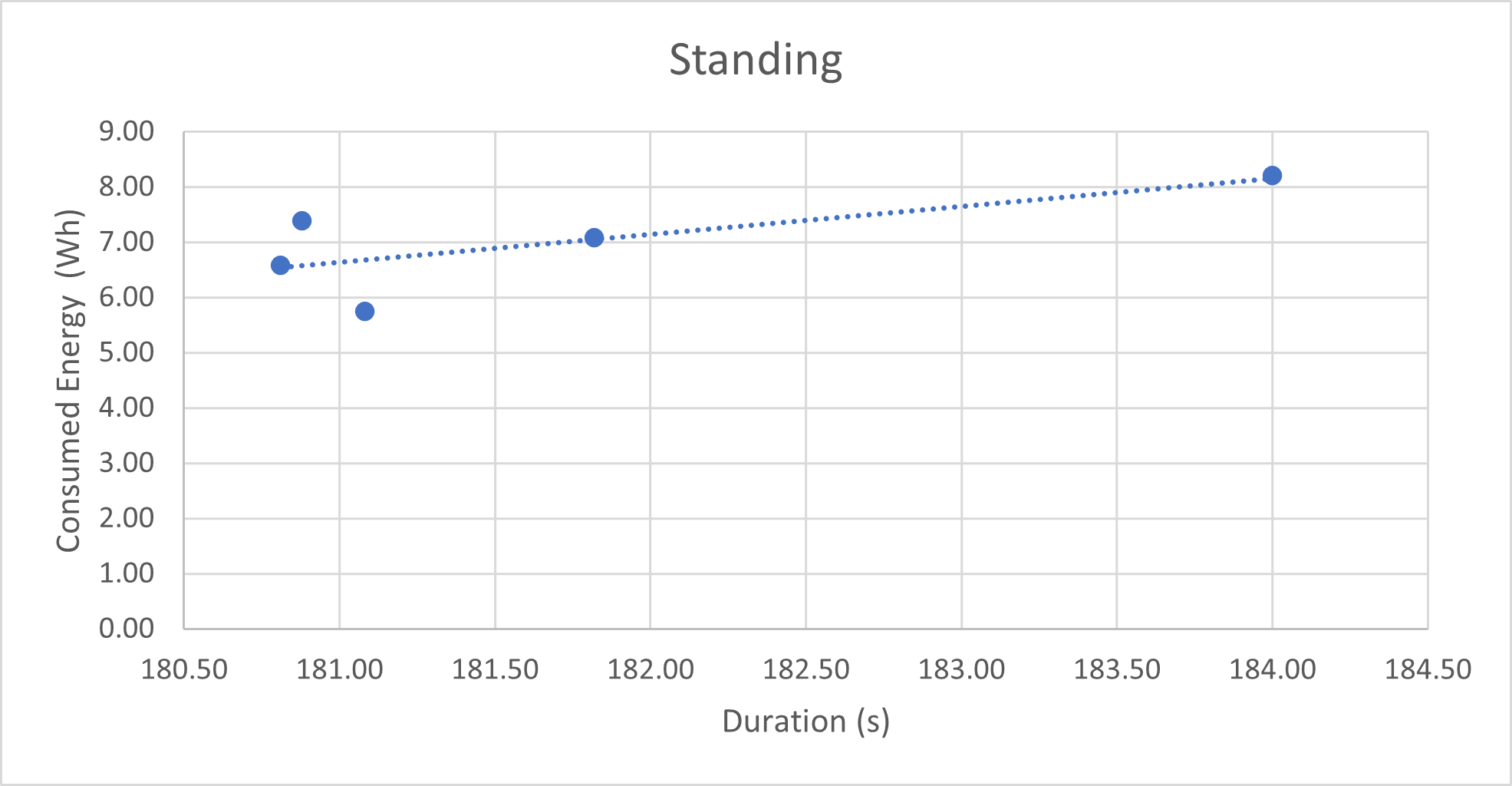}\label{fig:energy_standing}}}%
    \qquad
    \subfloat[\centering Energy consumption in forward walk at a constant speed of 0.4 m/s.]
    {{\includegraphics[width=0.80\columnwidth]{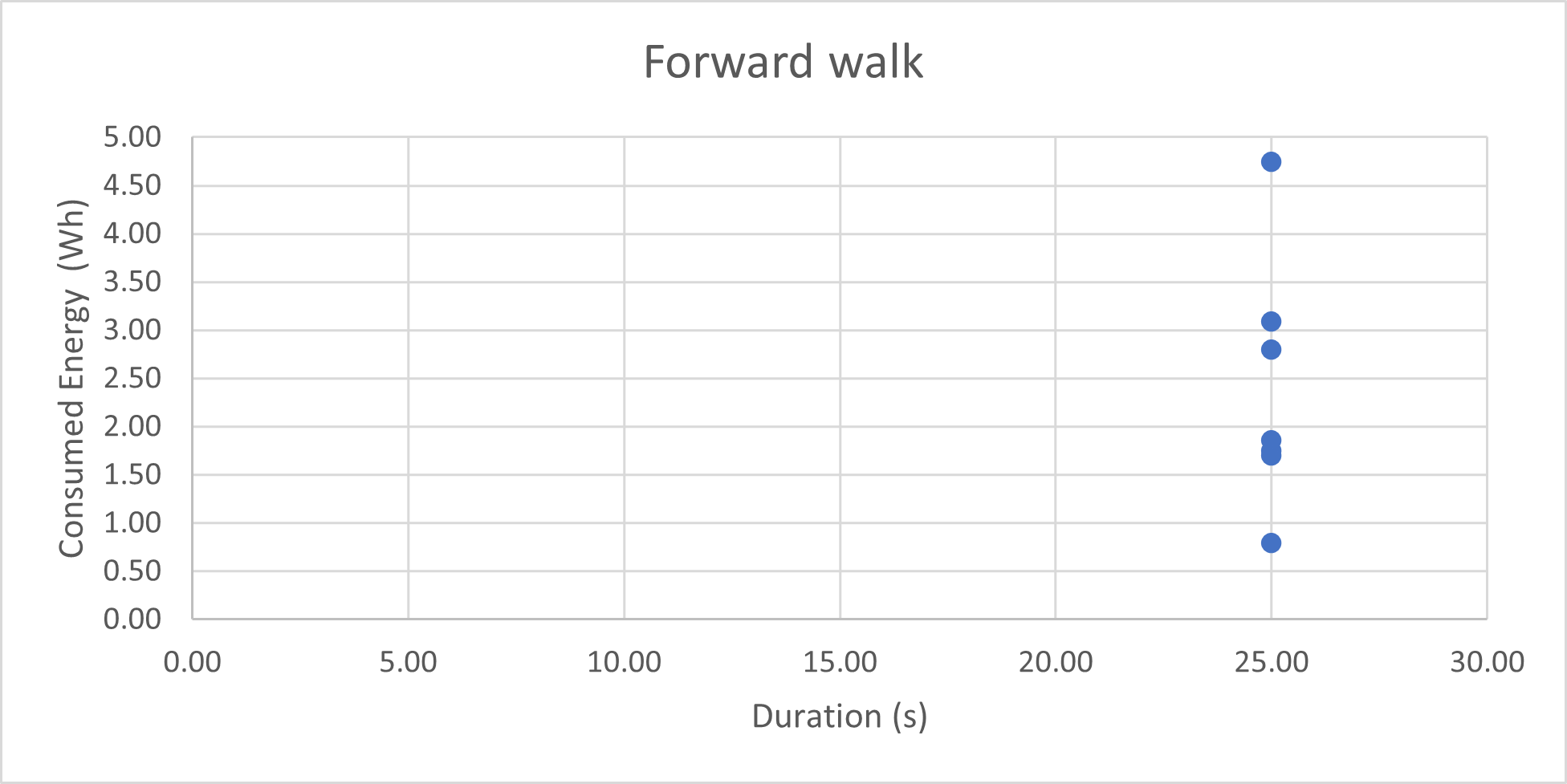}\label{fig:energy_forward_walk}}}%
    \qquad
    \caption{Energy consumption for various actions and their respective durations.
    }
    \label{fig:Econsumption}
\end{figure}

After the data collection was complete, the experimental values, including the state of charge and the voltage of each individual battery cell, were processed to fit into mathematical models. By applying \cref{eq:SoC} and multiplying the resulting value by the combined voltage for all battery cells, it was possible to estimate the remaining energy in the battery, expressed in Watt-hour (Wh). The energy consumption for a given action was then determined by comparing this value with the energy remaining from the previous iteration, subtracting the latter from the former. The graphics resulting from all measurements for the different actions are depicted in \cref{fig:Econsumption}. The energy consumption is lower when the MRP is in the lay down position, as this position allows the MRP's motors to be under the least strain, avoiding the need for balancing and movement actions (cf. \cref{fig:energy_layingdown}). Conversely, the most energy-intensive action is the forward walk movement at a constant speed, as it requires the displacement of the MRP, introducing greater operational demand on the motors. This movement also demonstrates the most variable energy consumption, as shown in \cref{fig:energy_forward_walk}, due to the continuous adjustments required to ensure the MRP's balance. The energy consumption for the remaining two actions is of the same order of magnitude, mainly due to the minor balance adjustments required by the MRP.

Considering the time measured for each iteration, the average power used during each action was computed by \cref{eq:average-power}. Additionally, taking into account the average power for each iteration, the overall average power was derived. The obtained values are presented in \cref{tab:Pmedia}. \cref{tab:Pmedia} can be used as a reference for estimating the energy consumption associated with complex actions, achieved by decomposing them into combinations of the four elementary measured in this research work. 
It is worth noting that the energy consumption associated with each action was determined by considering the action's duration, as expressed in \cref{eq:Econsumida}. The results, obtained for different action durations using \cref{eq:Econsumida} and taking into account the linear regressions derived from experimental data in \cref{fig:Econsumption}, are presented in \cref{tab:consumption_example}. It is important to note that the power consumption data was sampled over different durations for each activity.

\begin{equation}
    Average~Power =\frac{Energy~consumed}{\Delta t}
    \label{eq:average-power}
\end{equation}

\begin{equation}
    Consumed~Energy = \frac{Average Power \times \Delta t}{3600}
    \label{eq:Econsumida}
\end{equation}

\begin{table}[!ht]
    \centering
    \begin{tabular}{|m{1cm}|m{1cm}|m{1cm}|m{1cm}|m{1cm}|}
        \hline
        ~ \cellcolor{black}& Laying down&Crouching&Standing&Forward walk  \\
        \hline
        Average Power (W) & 62.99 & 150.14 & 138.58 & 344.65\\ 
        \hline
    \end{tabular}
    \caption{Average power spent for each action.}
    \label{tab:Pmedia}
\end{table}

\begin{table}
    \centering
    \begin{tabular}{|m{1cm}|m{1cm}|m{1cm}|m{1cm}|m{1cm}|}
    \hline
        ~ \cellcolor{black}& \multicolumn{3}{c}{~ ~ ~ ~ ~ Energy Consumed (Wh)} ~ & \\ \hline
        Duration (s) & Laying Down & Crouching & Standing & Forward Walk \\ \hline
        30 & 0.52 & 1.25 & 1.15 & 2.87 \\ \hline
        60 & 1.05 & 2.50 & 2.31 & 5.74 \\ \hline
        120 & 2.10 & 5.00 & 4.62 & 11.49 \\ \hline
        180 & 3.15 & 7.51 & 6.93 & 17.23 \\ \hline
        200 & 3.50 & 8.34 & 7.70 & 19.15 \\ \hline
        600 & 10.50 & 25.02 & 23.10 & 57.44 \\ \hline
    \end{tabular}
    \caption{Evolution of energy consumption over time for all actions.}
    \label{tab:consumption_example}
\end{table}

\section{Discussion \label{sec:Discussion}}
The experimental results presented in this paper allow assessing the power consumption for different actions of an MRP acting as a mobile communications cell. Assessing how each action influences overall energy usage and the endurance of an MRP is crucial in the context of wireless networks, particularly in dynamic networking scenarios that require the repositioning of the MRP over time.

The energy consumption characterization illustrates that energy consumption changes significantly with the type of action performed by the MRP. As shown in Table \ref{tab:consumption_example}, the energy spent during the forward walk was the highest, followed by crouching, standing, and laying down. It is worth noting that stationary actions, such as laying down and crouching, consume substantially less energy compared to dynamic actions such as forward walk.

The information on the average power spent during different actions can be used to optimize the MRP movement. In particular, the ability to extrapolate the energy usage for complex movements based on combinations of the four studied elementary actions allows defining targeted actions for an MRP to maximize the wireless network's availability while minimizing the MRP's energy consumption.

Our research work has some limitations that may be addressed in future work. First, the collection of a larger experimental dataset with more iterations for each type of action is worth considering. Additionally, given that not all complex MRP movements can be broken down into the four actions analyzed in this paper, additional measurements for other MRP actions may be performed, including evaluating varying speed values for the forward walk, introducing sideways movements, and considering movements across irregular terrains. 

The research work presented in this paper provides reference experimental values for characterizing the energy spent by MRPs. The results may be considered to formulate energy-aware MRP positioning approaches, in order to improve energy efficiency and endurance. Ultimately, this leads to enhanced availability of wireless networks enabled by MRPs.

\section{Conclusions\label{sec:Conclusions}}
A challenge when deploying mobile communications cells using MRPs is their availability. As MRPs are not permanently connected to the electrical grid, they rely on on-board batteries, which may quickly run out of power. As such, the characterization of MRP energy consumption is crucial to enable more energy-efficient MRP movements. 

In this paper, we present the experimental evaluation of the energy consumption of an MRP. The evaluation considered different actions performed by a real MRP, showing that energy consumption varies significantly with the type of action performed. By characterizing the energy requirements associated with different actions, MRP positioning can be optimized, extending its endurance and improving the availability of the wireless networks created.

For future work, it is worth enlarging the experimental dataset by increasing the number of iterations for each type of action and incorporating additional types of actions, such as sideways movements and movements over irregular terrains. This will pave the way for creating more accurate and generic energy consumption models.

\section*{Acknowledgments}
This work is financed by National Funds through the Portuguese funding agency, FCT -- Fundação para a Ciência e a Tecnologia, within project UIDB/50014/2020.

\printbibliography

@misc{UnitreeRobotics,
	title        = {{Unitree Go1}},
	author       = {Unitree},
	url          = {https://shop.unitree.com/products/unitreeyushutechnologydog-artificial-intelligence-companion-bionic-companion-intelligent-robot-go1-quadruped-robot-dog},
	note         = {Accessed on 06-07-2023}
}

@ARTICLE{Rodrigues2022_Joint-Energy-Performance,
  author={Rodrigues, Hugo and Coelho, André and Ricardo, Manuel and Campos, Rui},
  journal={IEEE Access}, 
  title={Joint Energy and Performance Aware Relay Positioning in Flying Networks}, 
  year={2022},
  volume={10},
  number={},
  pages={43848-43864},
  doi={10.1109/ACCESS.2022.3168695}
}

@Article{Romeo2020,
AUTHOR = {Romeo, Laura and Petitti, Antonio and Marani, Roberto and Milella, Annalisa},
TITLE = {Internet of Robotic Things in Smart Domains: Applications and Challenges},
JOURNAL = {Sensors},
VOLUME = {20},
YEAR = {2020},
NUMBER = {12},
ARTICLE-NUMBER = {3355},
URL = {https://www.mdpi.com/1424-8220/20/12/3355},
PubMedID = {32545700},
ISSN = {1424-8220},
}

@ARTICLE{Vermesan2020,
AUTHOR={Vermesan, Ovidiu and Bahr, Roy and Ottella, Marco and Serrano, Martin and Karlsen, Tore and Wahlstrøm, Terje and Sand, Hans Erik and Ashwathnarayan, Meghashyam and Gamba, Micaela Troglia},   
TITLE={Internet of Robotic Things Intelligent Connectivity and Platforms},    
JOURNAL={Frontiers in Robotics and AI},      
VOLUME={7},           
YEAR={2020},      
URL={https://www.frontiersin.org/articles/10.3389/frobt.2020.00104},     
DOI={10.3389/frobt.2020.00104},      
ISSN={2296-9144} 
}

@ARTICLE{Wang2023,
  author={Wang, Cheng-Xiang and You, Xiaohu and Gao, Xiqi and Zhu, Xiuming and Li, Zixin and Zhang, Chuan and Wang, Haiming and Huang, Yongming and Chen, Yunfei and Haas, Harald and Thompson, John S. and Larsson, Erik G. and Renzo, Marco Di and Tong, Wen and Zhu, Peiying and Shen, Xuemin and Poor, H. Vincent and Hanzo, Lajos},
  journal={IEEE Communications Surveys \& Tutorials}, 
  title={On the Road to 6G: Visions, Requirements, Key Technologies, and Testbeds}, 
  year={2023},
  volume={25},
  number={2},
  pages={905-974},
  doi={10.1109/COMST.2023.3249835}
}

@ARTICLE{Tataria2021,
  author={Tataria, Harsh and Shafi, Mansoor and Molisch, Andreas F. and Dohler, Mischa and Sjöland, Henrik and Tufvesson, Fredrik},
  journal={Proceedings of the IEEE}, 
  title={6G Wireless Systems: Vision, Requirements, Challenges, Insights, and Opportunities}, 
  year={2021},
  volume={109},
  number={7},
  pages={1166-1199},
  doi={10.1109/JPROC.2021.3061701}
}

@article{Liu2021,
  title = {A distributed deployment algorithm for communication coverage in wireless robotic networks},
  journal = {Journal of Network and Computer Applications},
  volume = {180},
  pages = {103019},
  year = {2021},
  issn = {1084-8045},
  doi = {https://doi.org/10.1016/j.jnca.2021.103019},
  url = {https://www.sciencedirect.com/science/article/pii/S1084804521000461},
  author = {Xiaojie Liu and Xingwei Wang and Jie Jia and Min Huang}
}

@article{Ludovica2023,
title = {ENSING: Energy saving based data transmission in Internet of Drones for 3D connectivity in 6G networks},
journal = {Ad Hoc Networks},
volume = {149},
pages = {103211},
year = {2023},
issn = {1570-8705},
doi = {https://doi.org/10.1016/j.adhoc.2023.103211},
author = {Ludovica {De Lucia} and Claudio Enrico Palazzi and Anna Maria Vegni}
}

\end{document}